\documentclass{ifacconf}
\usepackage{cite}
\usepackage{amsmath}
\usepackage{array}
\usepackage{graphicx}
\usepackage{natbib}
\usepackage{amssymb}
\usepackage{widetext}
\usepackage{multicol}
\usepackage{lipsum}  
\usepackage{url}

\newcommand{\bydef}{\stackrel{\Delta}{=}}

\newcommand{\beq}{\begin{equation}}
\newcommand{\eeq}{\end{equation}}
\newcommand{\beqa}{\begin{eqnarray}}
\newcommand{\eeqa}{\end{eqnarray}}
\newcommand{\beqan}{\begin{eqnarray*}}
\newcommand{\eeqan}{\end{eqnarray*}}
\newcommand{\bef}{\begin{figure}}
\newcommand{\enf}{\end{figure}}

\newcommand{\bi}{\begin{itemize}}
\newcommand{\ei}{\end{itemize}}
\newcommand{\bc}{\begin{center}}
\newcommand{\ec}{\end{center}}
\newcommand{\ba}{\begin{array}}
\newcommand{\ea}{\end{array}}
\newcommand{\be}{\begin{equation}}
\newcommand{\ee}{\end{equation}}
\newcommand{\beno}{\begin{equation*}}
\newcommand{\eeno}{\end{equation*}}
\newcommand{\beqna}{\begin{eqnarray}}
\newcommand{\eeqna}{\end{eqnarray}}
\newcommand{\bd}{\begin{displaymath}}
\newcommand{\ed}{\end{displaymath}}
\newcommand{\beqnd}{\begin{eqnarray*}}
\newcommand{\eeqnd}{\end{eqnarray*}}

\renewcommand{\ni}{\noindent}

\newcommand{\cqfd}{\hfill \rule{2mm}{2mm}\medbreak\indent}

\newtheorem{theorem}{\bf Theorem}[section]

\newtheorem{example}{\bf Example}[section]

\newtheorem{assumption}{\bf Assumption}

\usepackage{color}

\definecolor{red}{rgb}{1,0,0}
\newcommand{\red}{\textcolor{red}}

\definecolor{blu}{rgb}{0,0,1}

\definecolor{green}{rgb}{0,1,0}

\begin{document}
\begin{frontmatter}

\title{A practical method for the consistent identification of a module in a dynamical network}

\thanks{This work is supported
 by the Program Science Without Borders, CNPq - Conselho Nacional de Desenvolvimento Cient\'{\i}'fico e Tecnol—\'ogico, Brazil,  by the Belgian Programme on
Interuniversity Attraction Poles, initiated by the Belgian Federal
Science Policy Office, and by Wallonie-Bruxelles International.}

\author[michel]{Michel~Gevers} 
\author[alex]{Alexandre~Sanfelice~Bazanella} %
\author[alex]{Gian Vianna da Silva}
\address[michel]{ICTEAM, Louvain University, B1348 Louvain la Neuve, Belgium e-mail: Michel.Gevers@uclouvain.be}
\address[alex]{DELAE/UFRGS - Department of Automation and Energy,
        Universidade Federal do Rio Grande do Sul, Porto Alegre-RS, Brazil e-mail: bazanella@ufrgs.br; gianvs@live.com}


%

\begin{abstract}
We present a new and simple method for the identification of a single transfer function that is embedded in a dynamical network. In existing methods the  consistent identification of the desired transfer function relies on the positive definiteness of the spectral density matrix of the vector of all node signals, and it typically requires knowledge of the topology of the whole network. The positivity  condition is on the 
 internal signals and therefore can not be guaranteed a priori; in addition it is far from necessary. 
The new method of this paper provides simple 
conditions on which nodes to excite and which nodes to measure in order to produce a consistent estimate of the desired transfer function. Just as importantly,  it requires knowledge of the local topology only. 
\end{abstract}


\end{frontmatter}

\section{Introduction}

The present paper is  devoted to 
the identification of a particular transfer function (also called module or edge) within a network.
This problem is far from trivial because the interconnection structure
creates feedback loops, which may (and quite often do)
lead to the module of interest becoming unidentifiable from available signals.

A number of recent papers have addressed the problem of identification of a specific module  embedded within a dynamical network  
\cite{Vandenhof&Dankers&Heuberger&Bombois:13,Dankers&Vandenhof&Bombois&Heuberger:16,Gevers&Bazanella:15,Everitt&Bottegal&Hjalmarsson:17}. In \cite{Vandenhof&Dankers&Heuberger&Bombois:13,Dankers&Vandenhof&Bombois&Heuberger:16} the focus has been on the  identifiability of the desired transfer function, and on the question of which subset of node measurements will yield such identifiability. 
The  assumption is made in these papers  that the vector consisting
of all the node signals in the network is informative, and this assumption is crucial for the consistent estimation of the desired transfer function. 
This is a  strong persistence of  excitation condition, which is sufficient to identify
the whole network, but which  is  far from necessary for the identification of a single  module.
Moreover, it is an assumption on the internal signals rather than on the external excitation, which makes it  difficult to  enforce. 

The question as to how to guarantee that the vector of node signals is informative by appropriately choosing the externally applied signals is an experiment design problem. This problem was first approached in \cite{Gevers&Bazanella:15}, where it was illustrated on a simple 3-node network. It was shown in that paper that the choice of informative external signals that would lead to consistent identifiability of the desired transfer function depends both on the network topology and on the chosen identification method. A framework, based on \cite{Gevers&Bazanella&Bombois&Miskovic:09}, was proposed to determine which external signals need to be applied in order to make the vector of node signals informative. 

In \cite{Everitt&Bottegal&Hjalmarsson:17} a new approach has been proposed for the identification of a single module. Unlike the methods presented in \cite{Dankers&Vandenhof&Bombois&Heuberger:16} this new approach uses all external excitation signals that enter the network. Denoting $G_{ji}$ the desired transfer function to be identified, then \cite{Everitt&Bottegal&Hjalmarsson:17}  propose a two-step procedure, one of which consists of identifiying all transfers from all external signals to all the nodes that have a direct link to node $j$. 

All results cited above, that deal with the  identification of a single  module embedded within a dynamical network, 
require knowledge of the topology of the whole network. In this paper we present a completely different and very simple method which requires only local information about the topology. More precisely, if the module to be identified is $G_{ji}$, then the only topological information that is required for our new method is to know either to which nodes node $i$ connects (i.e. which nodes $k$ are such that $G_{ki}\neq 0$), or which nodes connect to $j$ (i.e. which nodes $k$ are such that $G_{jk}\neq 0$). 

Let us summarize this introduction as follows. The consistent identification of a specific  module $G_{ji}(q)$  contains both an experiment design aspect and a computational aspect. These  can be summarized in the following three problems:
\begin{itemize}
\item which nodes should be excited?
\item which nodes should be measured?
\item how to estimate $\hat G_{ji}(q)$ from these signals?
\end{itemize}
The new method proposed in this paper  solves all three problems at once. 
In addition, the method requires only local knowledge of the topology of the network, namely either the existence of the edges leaving node $i$, or the existence of the edges entering node $j$. 


The paper is organized as follows. The problem is stated in section~\ref{probstatement}. In section~\ref{background} we present the standard direct Prediction Error Method (PEM) that rewrites the network as a closed-loop system and then estimates the desired $G_{ji}$ by identifying all nonzero $G_{jk}$ that appear in the same row as $G_{ji}$. In section~\ref{casestudy} we apply this direct method to a 20-node network in order to illustrate the difficulty in arriving at excitation scenarii that yield a consistent estimate. We present the new method in section~\ref{new}, and we return to the case study in section~\ref{return} to illustrate its simplicity and effectiveness.

\section{Problem statement}\label{probstatement}

We adopt the network structure  of \cite{Gevers&Bazanella&Parraga:17},
in which the outputs of the nodes are denoted \linebreak $\{w_1(t), \ldots, w_L(t)\}$ and 
are related to each other and to the external excitation signals $r_j (t), j=1,\dots,L$
and the noise signals $v_j (t), j=1,\ldots,L$ by the following network equations:
\beqna 
\hspace{-2mm}\left[ \begin{array}{c}
w_1(t)\\ w_2(t)\\ \vdots \\ w_L(t) \end{array} \right] &=& \left[ \begin{array}{cccc}0 &  G_{12}^0(q) &  \ldots &\ 
G_{1L}^0 (q) \\G_{21}^0 (q)& 0   & \ddots &  G_{2L}^0 (q)\\Ê\vdots & \ddots &  \ddots &\vdots\\
G_{L1}^0(q) &  G_{L2}^0(q) &  \ldots &  0\end{array} \right]  \left[  \begin{array}{c}
w_1(t)\\ w_2(t)\\ \vdots \\ w_L(t) \end{array} \right] \nonumber \\
&&+ \left[ \begin{array}{c}  r_1 (t)\\  r_2(t) \\ \vdots \\  r_L(t)\end{array}  \right]  +
  \left[ \begin{array}{c} v_1(t) \\ v_2(t) \\ \vdots \\ v_L(t)\end{array} \right] \nonumber \\
&=& G^0(q) w(t) + r (t)+ v(t) .  \label{netmodel} 
\eeqna
where $q^{-1}$ is the delay operator and the superscript $^0$ denotes the real value of a quantity. 
The matrix $G^0(q)$ will be called the {\it  network matrix} and 
equation (\ref{netmodel}) the {\it network model}, and we will often omit the dependence on $t$ and on $q$ whenever 
it does not create ambiguity. We  assume that the network model has the properties specified in Assumption \ref{propriedades} below.

\begin{assumption}\label{propriedades}
The network model (\ref{netmodel}) has the following properties: 
\begin{enumerate}
\item all the transfer functions  $G_{ij}^0(q)$ are proper
\item there is a delay in every loop going from any $w_j(t)$ to itself
\item the stochastic processes $v_i(t)$ are stationary, zero mean and mutually independent: $E [ v_j (t) v_k(s) ] = 0~\forall t, s \in \Re$ for all $j\neq k$.
\item the external excitation signals $r_i(t)$ are quasi-stationary and uncorrelated with all noise signals $v_j(t)$ 
\item the network is internally stable.
\end{enumerate}
\end{assumption}

In this paper we  consider the problem of identifying  a particular transfer function,
say $G_{ji}^0(q)$, from measured node signals $w_k(t)$ and measured excitation signals $r_l(t)$.


\section{Background - The direct method}\label{background}


Perhaps the most natural approach towards the identification of a single transfer function
$G_{ji}$\footnote{From now on we omit the dependence on $q$ and $t$ whenever it creates no ambiguity.}
in a network is to estimate it from the scalar equation of (\ref{netmodel}) in which it appears,
based on the signals that appear in that equation. 
This corresponds to a closed-loop identification problem for a Multiple Input Single Output (MISO) feedback system with $L-1$ inputs and one output, namely $w_j$. 
Let us take, without loss of generality and to ease notation, $j=1$ and $i=2$, so that $G_{12}$ is the desired transfer function.

We  split up the vector $w$ into 
\be \label{wsplit}
w = \left[\begin{array}{c} w_1 \\ \tilde w_2\end{array}\right]
\ee
where $ \tilde w_2 \bydef \left[w_2 \ldots w_L \right]^T$. 
Correspondingly, we split up the matrix $G^0(q)$ into the $4$-block matrix
\be \label{4blockG}
G^0 = \left[\begin{array}{cc}0 & ( \begin{array}{ccc} G_{12}^0 &  \ldots &  G_{1L}^0 \end{array}) \\
\left( \begin{array}{c}G_{21}^0 \\ \vdots \\ G_{L1}^0 \end{array}\right) &
\left(\begin{array}{ccc}0 & \ddots & G_{2L}^0 \\ \ddots & \ddots & \vdots \\ G_{L2}^0 & \hdots & 0\end{array}\right)
\end{array}\right]
\ee
which we denote as
\be \label{4blockG2}
G^0 = \left[\begin{array}{cc}0 & \tilde G_1^0 \\ \tilde G_2^0 & \tilde G_3^0 \end{array}\right]
\ee
We can now rewrite the initial network model (\ref{netmodel}) as a  MISO feedback system as follows. First we rewrite (\ref{netmodel}) as
\be 
\left[\begin{array}{c}w_1 \\ \tilde w_2\end{array}\right]
=\left[\begin{array}{cc} 0 &  \tilde G_1^0\\ \tilde G_2^0& \tilde G_3^0\end{array}\right] \left[\begin{array}{c}w_1 \\ \tilde w_2\end{array}\right]  
   +   \left[\begin{array}{c} r_1 \\  \tilde r_2 \end{array}\right]   +   \left[\begin{array}{c} v_1 \\  \tilde v_2\end{array}\right] \label{MISO}
\ee


Next we rewrite (\ref{MISO}) in the traditional form of a MISO feedback system:  
\beqna
w_1& =&  \tilde G_1^0 \tilde w_2 +  r_1 +  v_1 \label{MISOFB1} \\
 &=& G_{12}^0w_2 + \sum_{k=3}^L G_{1k}^0 w_k +  r_1 +  v_1 \label{MISOFB2}\\
 \tilde w_2 &=& [I-\tilde G_3^0]^{-1} \{ \tilde G_2^0 w_1 
+   \tilde r_2 +   \tilde v_2 \} \nonumber \\
&&\label{MISOFB3} 
\eeqna

We observe that, if the objective is to identify $G_{12}^0(q)$, then it is natural to do so by identifying $\tilde G_1^0(q)$ (of which  $G_{12}^0(q)$ is the first element) 
directly from the scalar equation (\ref{MISOFB1}), by using the prediction error method.  
This problem setting is not the most realistic in large scale networks. 
Large scale networks tend to be highly sparse, meaning that most elements
of $\tilde G_1^0(q)$  are known to be zero. With this knowledge, 
then the identification can proceed as described before but identifying
only those elements that are known to be nonzero.

Let us define some notation in order to take advantage of 
the sparsity of the network. If (and only if) $G_{ik}$ is nonzero then node
$i$ is said to be an {\em out-neighbour} of node $k$; similarly, node $k$ is said to be an
{\em in-neighbour} of node $i$. Observe that the in-neighbours of a node $i$ correspond
to the nonzero elements of the $i$-th line of the network matrix $G^0$; similarly, the out-neighbours
of node $k$ correspond to the nonzero elements of the $k$-th column of $G^0$.
The set of in-neighbours of node $i$ is denoted $N_i^-$ and the set of 
out-neighbours of node $k$ is denoted $N_k^+$, their cardinalities being
$d_i^-$ and $d_k^+$, respectively. With these definitions
one can then write, in lieu of (\ref{MISOFB2}):
\begin{equation} \label{MISOFB4}
w_1 =  \sum_{k\in N_1^-} G_{1k}(q)w_k +  r_1 +  v_1
\end{equation}

Define further the vector containing the node signals of the in-neighbours 
of node 1, with dimension $d_1^-$:
$$
w_1^- = \left[\begin{array}{ccc} w_{l1} & \ldots & w_{ld_1^-}
\end{array}\right]^T
$$
where $l1, \ldots , ld_1^-$ are the indices corresponding to the in-neighbours of 
node $1$,
and the corresponding partition of the vector $\tilde G_1$ in (\ref{MISOFB1}):
$$
G_1^- (q) = \left[\begin{array}{ccc} G_{1, l1} (q) ~ \ldots ~ G_{1,ld_1^-} (q) \end{array}\right] .
$$
This is the vector of the nonzero transfer functions in (\ref{MISOFB2}).
Equation (\ref{MISOFB1}) can then be rewritten as 
\be \label{MISOFB5}
w_1 =  G_1^- (q) w_1^-  +  r_1 +  v_1.
\end{equation}
Next, let us define parametrized model structures $G_{1 k} (q, \theta)$ for each one of the transfer functions
in the vector $G_1^- (q)$, corresponding to a parametrized  $G_1^- (q,\theta)$,
and a model structure $H_1(q,\theta)$ for the noise $v_1 (t) = H_1^0 (q) e_1 (t)$ where $e_1(t)$ 
is a stationary zero-mean white noise process. Prediction error identification then proceeds by defining
the predictor:
\begin{equation}\label{predictor}
\hat w_1 (t,\theta ) =  H_1^{-1}(q,\theta) [ G_1^- (q,\theta ) w_1^- (t) +  r_1(t) ] 
\end{equation}
and minimizing the energy of the prediction error:
\begin{equation}\label{prediction_error}
\varepsilon (t, \theta) \bydefñ w_1(t) - \hat w_1(t,\theta).
\end{equation}

This direct application of prediction error identification to the network model has become known as
the {\em direct method} in recent literature. 
The following theorem from \cite{Vandenhof&Dankers&Heuberger&Bombois:13} gives sufficient conditions for the direct method to succeed
in providing consistent estimates. 

\begin{theorem}\label{teo_direto}
Consider a dynamic network (\ref{MISOFB1}) satisfying Assumption \ref{propriedades}
and the identification of the first row of the network matrix by the direct method described above.  
The estimate obtained for $G_1^- (q)$ is consistent if the following two conditions
are satisfied:
\begin{enumerate}
\item there exists $\theta^0$ such that $G_1^- (q,\theta) = (G_1^{-})^0 (q)$ and $H_1(q,\theta ) = H_1^0 (q)$
\item the spectral density of $w_1^-(t)$ is positive definite for almost all $\omega$.
\end{enumerate}
\end{theorem}

The problem  in the direct method, and also in other known methods as described in \cite{Vandenhof&Dankers&Heuberger&Bombois:13} is
condition 2 of the above theorem: it is a condition on internal signals $w_i(t)$. 
What is required for a proper experiment design are conditions
on the external signals  - $r_i(t)$ and $v_i(t)$ - that will enforce condition 2;
this issue is, however, not solved in the direct method\footnote{Actually it is mentioned as an open
question in the  last sentence of \cite{Vandenhof&Dankers&Heuberger&Bombois:13}.}.

This problem of transfer of excitation from the
external signals to the regressor used in the identification has been solved for SISO
systems in \cite{Gevers&Bazanella&Bombois&Miskovic:09}, where
necessary and sufficient conditions on the richness of the external excitation signals
for closed-loop systems have been given.
Although those tools can  also be used in the analysis of MISO problems,
as illustrated in \cite{Gevers&Bazanella:15}, no general results exist and, much
more importantly, the analysis requires knowledge of the whole network.


A case study in the next section illustrates the difficulties in this experiment design problem.

\section{A motivating case study}\label{casestudy}

\begin{figure}[h]
\renewcommand{\figurename}{Figure}
\hspace{-4.6cm}	\noindent\makebox[\textwidth]{\includegraphics[width=0.48\textwidth,height=0.4\textwidth]{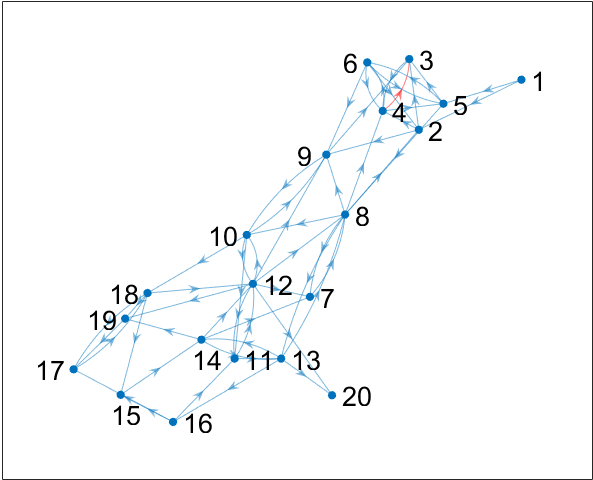}}
	\caption{The directed graph corresponding to the case study; the transfer function of interest is the red edge}
    \label{figGraph}
\end{figure}

Consider a network with $L=20$ nodes, of which we want to identify the transfer function
$G_{34}(q)$. The real network has the graph presented in Figure \ref{figGraph} and  the real transfer function 
is $G_{34}^0(q)=-0.3 q^{-1} +0.8q^{-2}$. All noises $v_i$ are white. The nonzero transfer functions are all of first or second order; the full $20\times 20$ matrix $G^0(q)$ is
given in Appendix \label{ap:G0}.

In order to apply the direct method, we need to know what are the in-neighbours of node $3$; these
are nodes $2$, $4$, $5$ and $9$. With these we form the regressor 
$$
w_3^- (t) = \left[\begin{array}{cccc} w_2(t) ~ w_4(t) ~ w_5(t) ~ w_9 (t) \end{array}\right]^T
$$

We take an Output-Error-like model structure, with
$H_3(q,\theta) = H_3(q) = 1$ and  full-order models for the transfer functions to be identified - 
the desired $G_{34}$ plus $G_{32}$, $G_{35}$ and $G_{39}$.
Then we apply the direct method with data collected from the real system under 18 different 
excitation scenarii: see Table~\ref{scenarii}. 
Each scenario consists of the excitation of a selected number of 
$r_i$ signals with a persistently exciting signal - stationary zero-mean white noise with unit variance. 
In each scenario, one thousand Monte-Carlo runs are performed, with ten thousand data in each run.
The average values obtained for the parameters in eighteen different excitation scenarii are given in Table \ref{averages}.
Recall that the real parameter values are $a_1^0=-0.3$ and $a_2^0 = 0.8$.

\begin{figure}[!ht]
	\renewcommand{\figurename}{Figure}
	{\includegraphics[width=0.52\textwidth,height=0.35\textwidth]{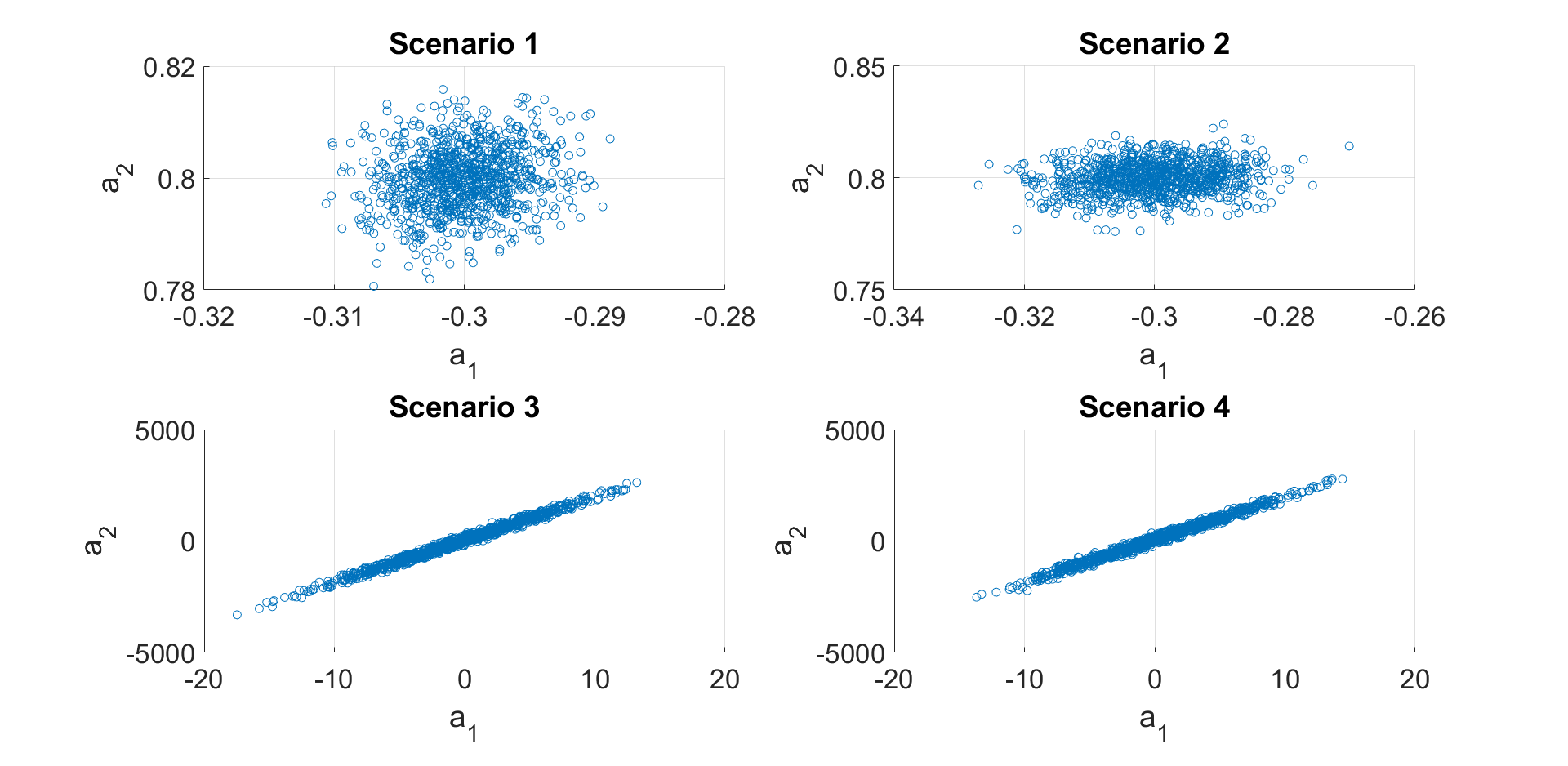}} 
	\caption{Scatter plot for Scenarios 1 to 4 (watch out: different scales for each plot); scenarii 1 and 2 provide consistent estimates, unlike scenarii 3 and 4}
    \label{MC_all_rs}
\end{figure}

In the first scenario, all $r_i$'s are excited, which guarantees that the regressor vector $w_3^-$ will have full-rank spectrum.
The result of the MC runs in the space of $G_{34}$ parameters is given in Figure \ref{MC_all_rs}. It is important to note that the scales of the top and bottom plots are widely different. 
It is seen from this Figure, as well as from Table \ref{averages}, that consistent estimation is obtained, as expected. 
But exciting all inputs in order to identify a single transfer function is obviously far from necessary and, most
importantly, far from feasible in a large network.

\begin{figure}[!ht]
	\renewcommand{\figurename}{Figure}
    \centering
	{\includegraphics[width=.52\textwidth,height=0.35\textwidth]{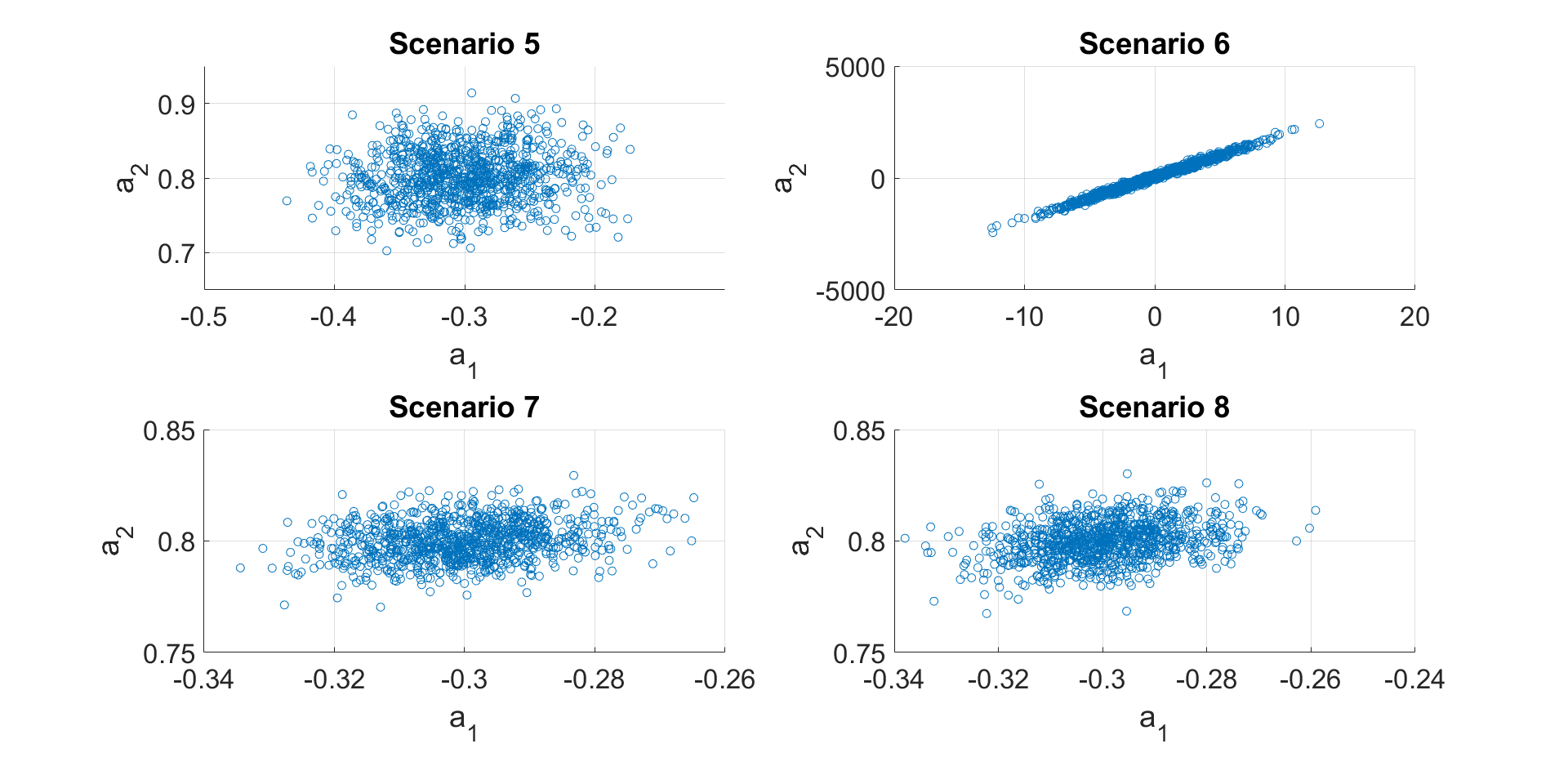}}
	\caption{Scatter plot for Scenarios 5 to 8 (watch out: different scales for each plot); exciting $r_5$ alone is enough to  provide consistent estimates}
    \label{MC_5_8}
\end{figure}

In fact, exciting a single input may be sufficient. We first try exciting only the node which is the
input to the desired transfer function, i.e. node 4, but it does not result in an informative experiment 
as can be seen both by the average values in Table \ref{averages} and in Figure \ref{MC_5_8}:  
consistency is not achieved and the variance tends to infinity. 
Exciting only $r_5$ (scenario 5 in Table \ref{averages}) proves to be enough, as can be seen in Figure \ref{MC_5_8},
but other ``closeby" excitation 
scenarii, i.e. exciting neighbors of the nodes 3 and 4  involved in the desired transfer function  $G_{34}$ do not work either.
On the other hand, exciting ``far away" nodes, like in scenario 17, does provide consistent estimation. 

\begin{table}[!ht]
\caption{Excitation Scenarios}
\label{scenarii}
\begin{tabular}{ll|lll}
Scenario & Inputs & Scenario & Inputs &  \\ \cline{1-4}
1 & All inputs ($r_1$ - $r_{20}$)        & 10 & $r_{2}$, $r_{6}$, $r_{7}$, $r_{8}$ &  \\
2 & $r_{3}$, $r_{4}$, $r_{5}$           & 11 & $r_{6}$, $r_{7}$, $r_{8}$, $r_{9}$ &  \\
3 & $r_{3}$                                & 12 & $r_{7}$, $r_{8}$, $r_{9}$, $r_{10}$ &  \\
4 & $r_{4}$                                & 13 & $r_{8}$, $r_{9}$, $r_{10}$, $r_{11}$ &  \\
5 & $r_{5}$                                & 14 & $r_{9}$, $r_{10}$, $r_{11}$, $r_{12}$ &  \\
6 & $r_{3}$, $r_{4}$                    & 15 & $r_{10}$, $r_{11}$, $r_{12}$, $r_{13}$ &  \\
7 & $r_{3}$, $r_{5}$                    & 16 & $r_{11}$, $r_{12}$, $r_{13}$, $r_{14}$ & \\
8 & $r_{4}$, $r_{5}$                    & 17 & $r_{1}$, $r_{7}$ & \\
9 & $r_{1}$, $r_{2}$, $r_{6}$, $r_{7}$ & 18 & $r_{1}$, $r_{16}$ & 
\end{tabular}
\end{table}

\begin{table}[!ht]
\caption{Parameters Estimation}
\label{averages}
\begin{tabular}{c|c|c|c|c|c|}
\cline{2-3} \cline{5-6}
                                & \multicolumn{2}{c|}{Parameters} &           & \multicolumn{2}{c|}{Parameters} \\ \hline
\multicolumn{1}{|c|}{Scenario} & $a_1$          & $a_2$          & Scenario & $a_1$          & $a_2$          \\ \hline
\multicolumn{1}{|c|}{1}         &    -0.3001     &     0.7998     & 10        &    -0.3004            &    0.8018             \\ \hline
\multicolumn{1}{|c|}{2}         &    -0.3002            &  0.8001              & 11        &    -0.2998            &    0.8073            \\ \hline
\multicolumn{1}{|c|}{3}         &       -0.3918         &     -17.8802           & 12        &    -0.3002            &     0.8387           \\ \hline
\multicolumn{1}{|c|}{4}         &     0.0526            &     70.9308           & 13        &     -0.2980           &     1.5186           \\ \hline
\multicolumn{1}{|c|}{5}         &      -0.3011          &    0.8020            & 14        &     -0.2981           &     1.7612           \\ \hline
\multicolumn{1}{|c|}{6}         &      -0.2124          &    15.5122            & 15        &      -0.3014          &      -0.7478          \\ \hline
\multicolumn{1}{|c|}{7}         &      -0.3003          &    0.8001            & 16        &      -0.3315          &      -12.0514          \\ \hline
\multicolumn{1}{|c|}{8}         &      -0.2997          &    0.8000            & 17        &     -0.3034           &     0.8007           \\ \hline
\multicolumn{1}{|c|}{9}         &      -0.3001         &    0.8007             & 18        &     -0.3080           &     0.7992           \\ \hline
\end{tabular}
\end{table}
In \cite{Gevers&Bazanella:15} we have shown that, to obtain an excitation scenario for the direct method that yields informative data, may require knowledge of the whole network even though only a single transfer function is to be identified.

\section{The New method}\label{new}


In this Section we  present a simple method which solves both the problem of transfer of excitation and that of measurement selection.
In other words, with this method we know a priori which inputs must be excited in order to obtain an
informative experiment and which nodes need be measured, using only local information about the network's topology.
%

The method is based on the input-output description of the system, which is 
obtained by rewriting it in a form that relates directly the external inputs $r$, the disturbances $v$
and the outputs $w$:
\beqna \label{iomodel}
w(t) &=& (I - G^0(q))^{-1}[r(t) + v(t)] \\
&=& T^0(q) r(t) + \bar v(t) \label{iomodel2}
\eeqna
where \be \label{iorelation}
T^0(q) \bydef (I - G^0(q))^{-1}, ~~~\bar v(t) \bydef (I - G^0(q))^{-1}v(t).
\ee

The description (\ref{iomodel2}) will be called the {\it input-output (I/O) model} of the network. It is well known that one can obtain a consistent estimate $\hat T(q)$ of $T^0(q)$ from $\{w,r\}$ data; this is an open loop MIMO identification problem. 

Suppose we have an estimate $\hat T(q)$ of $T^0$; then an estimate $\hat G$ of $G^0$ can be obtained by solving either one of the following two equations for $\hat G(q)$, each one being a set of $L^2$ linear equations:
\begin{eqnarray} 
&& \hat T(q)  (I - \hat G(q)) = I  \label{TzeroT2} \\
&&  (I - \hat G(q)) \hat T(q)  = I  \label{TzeroT3} .
\end{eqnarray}
If only  one particular transfer function in $G^0$ 
is desired, then it can be obtained by solving a subset of these equations; this rationale is the basis of our
method. 
To describe the method, we  introduce some notations that are specific to the identification of the transfer function $G_{ji}$. For simplicity, we also refer to the identification of an edge of the network to refer to the identification of its transfer function. \\
{\bf Notations:} \\ 
$\bullet$ $G_{N_i^+,i} (q)=$ column vector of the out-going edges  of $i$ \\
$\bullet$ $T_{j, i} (q)=$ the $j, i $ element of matrix $T(q)$ \\
$\bullet$ $T_{N_i^+ L} (q)=$ submatrix of $T (q)$ made up of its rows in $N_i^+$\\
$\bullet$ $T_{N_i^+,i} (q)= i$-th column of $T_{N_i^+,L}(q)$  \\
$\bullet$ $T_{N_i^+N_i^+} (q) =$ submatrix of $T(q)$ made up of its rows and columns in $N_i^+$. \\
$\bullet$ $E_i = i$-th column of the identity matrix.\\

Consider the network (\ref{netmodel}) and assume that it is desired to only identify a specific transfer function $G_{ji}$. 

\begin{theorem}\label{theo:module1}
Perform an experiment under the following conditions:
\begin{itemize}
\item[-] excite  node $i$ and all its $N_i^+$ out-neighbors with sufficiently rich signals
\item[-] measure the node signals at the $N_i^+$ out-neighbors of node $i$.
\end{itemize}
Under these experimental conditions and using full-order models for the elements of the matrix $T^0$, 
consistent estimates $\hat{T}_{N_i^+N_i^+}$ and $\hat{T}_{N_i^+,i}$ of $T_{N_i^+N_i^+}^0$ and $T_{N_i^+,i}^0$
can be obtained by standard open-loop MIMO identification.
From these, a consistent estimate of $G^0_{N_i^+,i}$ is obtained by 
\be \label{solGSi}
\hat{G}_{N_i^+,i} (q) = [\hat{T}_{N_i^+N_i^+} (q)]^{-1} \hat{T}_{N_i^+,i}(q)
\ee
\end{theorem}
\ni {\bf Proof:} First note that identification of $T_{N_i^+N_i^+}^0$ and $T_{N_i^+,i}^0$ is an open-loop identification problem which,
under the specified experimental conditions and with full order models for the elements of these matrices, 
provides consistent estimates. Now, consider the system of equations
\begin{equation}\label{hatTzeroT2}
T(q)  (I - G(q)) = I, 
\end{equation}
where it is desired to compute $G_{ji}(q)$ from $T(q)$. 
The  desired $G_{ji}$ appears only in a subset of these equations,
its computation resting entirely on the solution of the $i$-th column of (\ref{hatTzeroT2}).
In the $i$-th column of $I-G$, the only nonzero elements are: the desired $G_{ji}$, the $1$ at position $(i,i)$, and the $G_{ki}$ corresponding to the
remaining $d_i^+ $ out-neighbors  of $i$. As a result, the columns of $T$ corresponding to the zero elements of the $i$-th column
of $G$ do not contribute to the computation of $G_{ji}$ using (\ref{hatTzeroT2}). It follows that these columns need not be identified and,
therefore, the corresponding $r_k$ are not required. Stated otherwise, for the identification of $G^0_{ji}$, it is only required to excite
node $i$ and its  $d_i^+$ out-neighbors.\\
To compute $G_{ji}$ we compute the $i$-th column of $(I-G)$, of which $G_{ji}$ is an element. From (\ref{hatTzeroT2}), we thus get 
\be \label{Teqn1}
T(I-G)_{:i} = E_i.
\ee
Now let $C$ be a selector matrix that selects the rows of $T$ that are in the set
$N_i^+$ of out-neighbors of $i$. Premultiplying (\ref{Teqn1}) by this $C$ yields
\be \label{Teqn2}
CT(I-G)_{:i} = C_{:i} =0~~\Leftrightarrow ~~T_{N_i^+,L} (I-G)_{:i} = 0
\ee
because the $i$-th column of $C$ is zero. Since $(I-G)_{:i}$ contains a $1$ in position  $i$ and nonzero elements only
in the positions corresponding to $N_i^+$, the last equation is equivalent with 
\be \label{Teqn3}
T_{N_i^+,i} -  T_{N_i^+N_i^+} G_{N_i^+,i} = 0,
\ee
from which the result (\ref{solGSi}) follows. 
That $T_{N_i^+N_i^+}$ is nonsingular follows from Proposition 5.1 in
\cite{Hendrickx&Gevers&Bazanella:17}\footnote{Just apply it to ${\cal A} = {\cal C}= N_i^+$.}.
\cqfd
The desired transfer function $G_{ji}$ is an element of $G_{N_i^+,i}$; hence Theorem~\ref{theo:module1} 
 provides a new method for the identification of a single embedded module.
The method rests on the identification of a submatrix of the transfer function matrix $T^0$; the number of elements
$T^0_{kl}$ that need to be identified is $d_i^+ \times (1 +d_i^+)$, where $d_i^+$ is the number of out-neighbors
of the input node $i$ of the desired $G_{ji}$. What is most important is that the theorem completely solves both  the
informativity and the identifiability questions for the identification of a single embedded module, namely  which nodes
need to be excited and which nodes need to be measured, in addition to providing a computational method for the estimation of $G_{ji}$.  Moreover, this solution requires only local information
about the network's topology, namely what nodes are the out-neighbors of node $i$.

In the same spirit, we can derive a dual method by manipulating equation (\ref{TzeroT3}) instead of (\ref{TzeroT2}).
In this case only the $j$-th  equation is relevant and we get:
\be \label{TTeqn11}
(I-G)_{j:} T = E_j^T
\ee
Taking a selector matrix $C$ that selects the columns of $T$ that correspond to the in-neighbors of $j$ yields:
\be \label{TTeqn22}
(I-G)_{j:} TC = C_{j:} =0~~\Leftrightarrow ~~(I-G)_{j:} T_{L, N_j^-} = 0
\ee
because the $j$-th line of $C$ is zero. Since $(I-G)_{j:}$ contains a $1$ in position $j$ and nonzero elements only
in the positions corresponding to $N_j^-$, the last equation is equivalent with 
\be \label{Teqn3}
T_{j,N_j^-} -  G_{j,N_j^-} T_{N_j^-N_j^-}  = 0,
\ee
and we have proven the following result. 

\begin{theorem}\label{theo:module2}
Perform an experiment under the following conditions:
\begin{itemize}
\item[-] excite the node signals at all $N_j^-$  in-neighbors of node $j$ with sufficiently rich signals
\item[-] measure node $j$ and the node signals of  its $N_j^-$ in-neighbors.
\end{itemize}
Under these experimental conditions and using full-order models for the elements of the matrix $T^0$,
consistent estimates $\hat{T}_{N_j^-N_j^-}$ and $\hat{T}_{j,N_j^-}$ of $T_{N_j^-N_j^-}^0$ and $T_{j,N_j^-}^0$
can be obtained by standard open-loop MIMO identification.
From these estimates, a consistent estimate of $G_{j,N_j^-}$ is given by 
\be \label{solGSi2}
\hat{G}_{j,N_j^-} (q) =  \hat{T}_{j,N_j^-}(q) [\hat{T}_{N_j^-N_j^-} (q)]^{-1}
\ee
The desired transfer function $G_{ji}$ is an element of $G_{j,N_j^-}$.
\ni {\bf Proof:} The proof is the exact dual of the proof of Theorem~\ref{theo:module1}, and is therefore omitted.
\cqfd
\end{theorem}
\vspace{-5mm}

The desired transfer function $G_{ji}$ is an element of $G_{j,N_j^-}$. The identification  of $G^0_{ji}$ rests on the identification of a submatrix of $T^0$ that contains $(d_j^- +1)\times d_j^-$ elements, where $d_j^-$ is the number of in-neighbors of the output node $j$ of the desired $G_{ji}$. Again, the identification of $G_{ji}$ using the dual method of Theorem~\ref{theo:module2} requires only local information about the network's topology, namely what nodes are the in-neighbors of node $j$.

\ni {\bf Comment.} If local information is available about both the out-neighbors of node $i$ and the in-neighbors of node $j$, then the user has a choice, for the identification of $G_{ji}$, of applying either Theorem~\ref{theo:module1} or Theorem~\ref{theo:module2}. The decision may depend on the practically available excitation and measurement scenarii. If both scenarii are possible, he/she will then obviously chose to apply the method that requires the smallest number of transfer functions $T^0_{kl}$ to be identified, by applying Theorem~\ref{theo:module1} if $d_i^+ \leq d_j^-$ and Theorem~\ref{theo:module2} otherwise.

\section{The case study revisited}\label{return}

Let us now apply our  method  to the identification of $G_{34}(q)$ in our 20-node case study of section~\ref{casestudy}. 
It is seen in the graph that node $4$ has 3 out-neighbors, i.e. $d_4^+ = 3$, while node $3$ has 4 in-neighbors, i.e. $d_3^-=4$. Thus we identify $G_{34}(q)$ using the method of Theorem \ref{theo:module1}, noting that 
the set of  outneighbors of node $4$  is $N_4^+ = \{ 3, 5, 6 \}$. So we need an experiment in which
we measure these three nodes and excite them plus node $4$. We have excited each 
input $r_i, ~i = 3, 4, 5, 6$ with independent white noises with unit variance, and also applied unmeasured
noise signals $v_i, ~i = 3, 4, 5, 6$ with variance $10^{-6}$.

From this experiment  we have identified the nine  transfer functions in the matrix $T_{N_4^+,N_4^+}$ and the three
transfer functions in the vector $T_{N_4^+,4}$. We have performed black-box identification
of order six models for these twelve transfer functions by the instrumental variable method in MatLab's
identification toolbox.  The actual transfer functions $T^0_{kl}$ are of very large order, but models of order six were enough
to get  a fit above $99\%$. 
The identified transfer functions are in the form
$$
T_{ij} = \frac{b_0+b_1z^{-1}+b_2z^{-2}+b_3z^{-3}+b_4z^{-4}+b_5z^{-5}+b_6z^{-6}}{1+a_1z^{-1}+a_2z^{-2}+a_3z^{-3}+a_4z^{-4}+a_5z^{-5}+a_6z^{-6}} \\
$$
and their identified parameters are given below, with eight significant digits.

Some of the estimated transfer functions had unstable states that were removed. This was accomplished by determining the balanced state space realization of the unstable transfer function with the \texttt{dbalreal} command from MATLAB, which also returns the vector containing the diagonal of the observability/controllability matrix. States that the corresponded to \texttt{Inf} values on this vector were removed with the command \texttt{dmodred}.

\begin{eqnarray*}
&& \mathbf{T_{33}} \\
a_1 & = &  \phantom{-} 3.8481674 \times 10^{-2} \\
a_2 & = &  \phantom{-} 3.0797313 \times 10^{-1} \\
a_3 & = & -6.6201436 \times 10^{-1} \\
a_4 & = &  \phantom{-} 1.7938664 \times 10^{-2} \\
a_5 & = &  \phantom{-} 3.0075344 \times 10^{-2} \\
a_6 & = & \phantom{-} 0 \\
b_0 & = &  \phantom{-} 1.0005030 \\
b_1 & = &  \phantom{-} 3.1534410 \times 10^{-2} \\
b_2 & = & -1.7410433 \times 10^{-2} \\
b_3 & = & -2.2352584 \times 10^{-2} \\
b_4 & = & -2.0168467 \times 10^{-2} \\
b_5 & = &  \phantom{-} 8.5201918 \times 10^{-7} \\
b_6 & = & \phantom{-} 0
\end{eqnarray*}
\begin{eqnarray*}
&& \mathbf{T_{34}} \\
a_1 & = & -4.6166469 \times 10^{-1} \\
a_2 & = &  \phantom{-} 3.1515405 \times 10^{-1} \\
a_3 & = & -8.3825398 \times 10^{-1} \\
a_4 & = &  \phantom{-} 3.5551666 \times 10^{-1} \\
a_5 & = & -5.3388080 \times 10^{-3} \\
a_6 & = & \phantom{-} 0 \\
b_0 & = & -4.5876773 \times 10^{-3} \\
b_1 & = & -3.2505403 \times 10^{-1} \\
b_2 & = &  \phantom{-} 8.0327554 \times 10^{-1} \\
b_3 & = & -3.6555712 \times 10^{-1} \\
b_4 & = &  \phantom{-} 1.3053090 \times 10^{-2} \\
b_5 & = &  \phantom{-} 1.2556216 \times 10^{-6} \\
b_6 & = & \phantom{-} 0
\end{eqnarray*}
\begin{eqnarray*}
&& \mathbf{T_{35}} \\
a_1 & = &  \phantom{-} 2.4570327 \times 10^{-1} \\
a_2 & = &  \phantom{-} 4.5393517 \times 10^{-1} \\
a_3 & = & -4.7766990 \times 10^{-1} \\
a_4 & = & -5.0273260 \times 10^{-2} \\
a_5 & = & -1.0096247 \times 10^{-2} \\
a_6 & = & -5.3395820 \times 10^{-2} \\
b_0 & = & -7.8937977 \times 10^{-2} \\
b_1 & = & -2.8932660 \times 10^{-1} \\
b_2 & = & -2.2567717 \times 10^{-1} \\
b_3 & = & -1.1488447 \times 10^{-1} \\
b_4 & = & -4.0703106 \times 10^{-2} \\
b_5 & = &  \phantom{-} 5.8362039 \times 10^{-3} \\
b_6 & = & -1.7599509 \times 10^{-17} 
\end{eqnarray*}
\begin{eqnarray*}
&& \mathbf{T_{36}} \\
a_1 & = &  \phantom{-} 3.2094476 \times 10^{-1} \\
a_2 & = &  \phantom{-} 3.8846671 \times 10^{-1} \\
a_3 & = & -5.6921619 \times 10^{-1} \\
a_4 & = & -1.3026078 \times 10^{-1} \\
a_5 & = &  \phantom{-} 2.9584499 \times 10^{-3} \\
a_6 & = &  \phantom{-} 2.2717961 \times 10^{-2} \\
b_0 & = &  \phantom{-} 1.4074532 \times 10^{-1} \\
b_1 & = & -2.1822255 \times 10^{-1} \\
b_2 & = &  \phantom{-} 1.0131479 \times 10^{-1} \\
b_3 & = & -1.0562591 \times 10^{-1} \\
b_4 & = &  \phantom{-} 1.7462396 \times 10^{-2} \\
b_5 & = &  \phantom{-} 9.7104478 \times 10^{-3} \\
b_6 & = &  \phantom{-} 1.7215627 \times 10^{-18} 
\end{eqnarray*}
\begin{eqnarray*}
&& \mathbf{T_{53}} \\
a_1 & = & -2.1525075 \times 10^{-2} \\
a_2 & = &  \phantom{-} 4.1218212 \times 10^{-1} \\
a_3 & = & -6.7085915 \times 10^{-1} \\
a_4 & = &  \phantom{-} 9.8264186 \times 10^{-2} \\
a_5 & = & -3.8151595 \times 10^{-2} \\
a_6 & = & \phantom{-} 0 \\
b_0 & = & -2.4299859 \times 10^{-4} \\
b_1 & = &  \phantom{-} 4.2864147 \times 10^{-4} \\
b_2 & = &  \phantom{-} 4.9627613 \times 10^{-1} \\
b_3 & = & -2.0867406 \times 10^{-2} \\
b_4 & = &  \phantom{-} 2.1384076 \times 10^{-2} \\
b_5 & = & -8.0865862 \times 10^{-6} \\
b_6 & = & \phantom{-} 0 
\end{eqnarray*}
\begin{eqnarray*}
&& \mathbf{T_{54}} \\
a_1 & = & -1.9833111 \times 10^{-1} \\
a_2 & = &  \phantom{-} 1.8199887 \times 10^{-1} \\
a_3 & = & -7.5568457 \times 10^{-1} \\
a_4 & = &  \phantom{-} 1.3184591 \times 10^{-1} \\
a_5 & = &  \phantom{-} 9.6356440 \times 10^{-2} \\
a_6 & = & \phantom{-} 0 \\
b_0 & = & -5.5988128 \times 10^{-4} \\
b_1 & = &  \phantom{-} 4.9701271 \times 10^{-1} \\
b_2 & = & -1.1035300 \times 10^{-1} \\
b_3 & = & -9.1997426 \times 10^{-2} \\
b_4 & = & -7.2354865 \times 10^{-3} \\
b_5 & = & -3.5608761 \times 10^{-5} \\
b_6 & = & \phantom{-} 0 
\end{eqnarray*}
\begin{eqnarray*}
&& \mathbf{T_{55}} \\
a_1 & = & -2.4661243 \times 10^{-1} \\
a_2 & = &  \phantom{-} 7.7627922 \times 10^{-1} \\
a_3 & = & -7.3814227 \times 10^{-1} \\
a_4 & = &  \phantom{-} 3.9799183 \times 10^{-1} \\
a_5 & = & -2.7861447 \times 10^{-1} \\
a_6 & = &  \phantom{-} 2.4048797 \times 10^{-2} \\
b_0 & = &  \phantom{-} 1.0017417 \\
b_1 & = & -2.7825384 \times 10^{-1} \\
b_2 & = &  \phantom{-} 7.0393709 \times 10^{-1} \\
b_3 & = & -8.6350269 \times 10^{-1} \\
b_4 & = &  \phantom{-} 3.4837217 \times 10^{-1} \\
b_5 & = & -3.0949202 \times 10^{-1} \\
b_6 & = & -1.1725324 \times 10^{-17} 
\end{eqnarray*}
\begin{eqnarray*}
&& \mathbf{T_{56}} \\
a_1 & = &  \phantom{-} 6.2495573 \times 10^{-2} \\
a_2 & = &  \phantom{-} 4.2084437 \times 10^{-1} \\
a_3 & = & -6.0218364 \times 10^{-1} \\
a_4 & = &  \phantom{-} 7.4963364 \times 10^{-2} \\
a_5 & = & \phantom{-} 0 \\
a_6 & = & \phantom{-} 0 \\
b_0 & = &  \phantom{-} 1.8924638 \times 10^{-2} \\
b_1 & = &  \phantom{-} 8.5600937 \times 10^{-2} \\
b_2 & = & -2.2788412 \times 10^{-2} \\
b_3 & = & -5.6722905 \times 10^{-2} \\
b_4 & = & -3.8189237 \times 10^{-6} \\
b_5 & = & \phantom{-} 0 \\
b_6 & = & \phantom{-} 0 
\end{eqnarray*}
\begin{eqnarray*}
&& \mathbf{T_{63}} \\
a_1 & = &  \phantom{-} 4.8106647 \times 10^{-1} \\
a_2 & = & -7.2321028 \times 10^{-2} \\
a_3 & = & -5.1645253 \times 10^{-1} \\
a_4 & = & -4.1069797 \times 10^{-1} \\
a_5 & = &  \phantom{-} 3.1824379 \times 10^{-1} \\
a_6 & = & -9.2033189 \times 10^{-3} \\
b_0 & = & -8.8116822 \times 10^{-4} \\
b_1 & = & -3.8823253 \times 10^{-2} \\
b_2 & = & -1.8354666 \times 10^{-2} \\
b_3 & = &  \phantom{-} 2.6939801 \times 10^{-2} \\
b_4 & = &  \phantom{-} 4.8460620 \times 10^{-3} \\
b_5 & = & -5.1528647 \times 10^{-3} \\
b_6 & = & -1.0676972 \times 10^{-18} 
\end{eqnarray*}
\begin{eqnarray*}
&& \mathbf{T_{64}} \\
a_1 & = & -1.0520487 \\
a_2 & = &  \phantom{-} 1.0895993 \\
a_3 & = &  \phantom{-} 1.0072416 \\
a_4 & = &  \phantom{-} 1.0194335 \\
a_5 & = & -5.5836734 \times 10^{-1} \\
a_6 & = & \phantom{-} 0 \\
b_0 & = & -3.7758880 \times 10^{-2} \\
b_1 & = &  \phantom{-} 3.8087588 \times 10^{-2} \\
b_2 & = & -1.7630291 \times 10^{-2} \\
b_3 & = & -1.3593510 \times 10^{-2} \\
b_4 & = &  \phantom{-} 1.0967873 \times 10^{-2} \\
b_5 & = & -2.0085089 \times 10^{-5} \\
b_6 & = & \phantom{-} 0
\end{eqnarray*}
\begin{eqnarray*}
&& \mathbf{T_{65}} \\
a_1 & = &  \phantom{-} 8.5049286 \times 10^{-1} \\
a_2 & = &  \phantom{-} 3.7655852 \times 10^{-1} \\
a_3 & = & -4.0517460 \times 10^{-1} \\
a_4 & = & -5.2111539 \times 10^{-1} \\
a_5 & = &  \phantom{-} 1.9891282 \times 10^{-2} \\
a_6 & = & -1.8172786 \times 10^{-2} \\
b_0 & = & -4.7802708 \times 10^{-2} \\
b_1 & = & -1.4941565 \times 10^{-2} \\
b_2 & = &  \phantom{-} 1.2826575 \times 10^{-2} \\
b_3 & = &  \phantom{-} 3.8958096 \times 10^{-2} \\
b_4 & = &  \phantom{-} 1.6370716 \times 10^{-2} \\
b_5 & = & -1.6243042 \times 10^{-2} \\
b_6 & = & -1.9197463 \times 10^{-18}  
\end{eqnarray*}
\begin{eqnarray*}
&& \mathbf{T_{66}} \\
a_1 & = &  \phantom{-} 7.8166080 \times 10^{-1} \\
a_2 & = &  \phantom{-} 9.1263291 \times 10^{-1} \\
a_3 & = & -2.3568030 \times 10^{-3} \\
a_4 & = & \phantom{-} 0 \\
a_5 & = & \phantom{-} 0 \\
a_6 & = & \phantom{-} 0 \\
b_0 & = &  \phantom{-} 9.9342966 \times 10^{-1} \\
b_1 & = &  \phantom{-} 7.7673764 \times 10^{-1} \\
b_2 & = &  \phantom{-} 9.1624167 \times 10^{-1} \\
b_3 & = &  \phantom{-} 6.6391065 \times 10^{-6} \\
b_4 & = & \phantom{-} 0 \\
b_5 & = & \phantom{-} 0 \\
b_6 & = & \phantom{-} 0 \\
\end{eqnarray*}

With these transfer function estimates, we then computed the estimate
$\hat{G}_{N_4^+,4}$  from equation (\ref{Teqn3}), of which the desired transfer function $G_{34}$
is one of the elements. In so doing, a transfer function of order $62$ was obtained for $\hat G_{34}$. We then proceeded with $H_2$ 
order reduction using the knowledge about the model class of the actual  transfer function, which is
of the form $G_{34}(q,\theta) = a_1 q^{-1} + a_2 q^{-2}$. 
The parameters of $\hat{G}_{34}(q,\theta)$ were estimated by calculating $\hat{G}_{N_4^+,4}$ through the Equation (\ref{solGSi}), on 100 frequency points between $e^{j0}$ and $e^{j2\pi}$, then $\hat{a}_1$ and $\hat{a}_2$ were estimated via the least squares method.
The following estimates were obtained: $\hat{a}_1 = -0.2992$ and
$\hat{a}_2 = 0.7979$, which confirms the efficiency of our new method. 
All details about this case study will be provided in the ArXiv version of this paper.

\section{Conclusions}\label{conclu}

The direct method for the identification of a module in a network is well known and easy to apply
if informative data are available. However, as we have illustrated through a 20-node example,
there is no practical way to design an informative experiment for this method. 
Though the excitation requirements vary for different methods and for different 
input selections, they all suffer from this  limitation.

We have presented an identification method for which the design of informative experiments
is obviated. It consists in performing the identification of part of the network's input-output model and then
recovering the desired module from these identified transfer functions. Because identification of the I-O model
is an open loop identification problem, it is clear which are the inputs that must be excited and
the critical issue becomes to determine what is the smallest set of I-O transfer functions that  must
be estimated in order to be able to uniquely recover the desired module. 
We have shown that this smallest set depends strictly on the network's local topology - that is, on
what are the neighbours of nodes $i$ and $j$ - and we have provided two choices for it, one
involving the in-neighbors of the end node of the desired module, and another one the out-neighbors of its source node.  

We have illustrated our method by a successful application to the 20-node example.


\bibliography{migrefs}

\appendix
\section{The 20x20 Matrix}
The matrix $G^0(q)$:

\begin{equation*}
\label{eqG}
G^0(q)=
\begin{bmatrix}
G_1 & G_2\\ 
G_3 & G_4
\end{bmatrix}
\end{equation*}

where 

\begin{equation*}
\label{eqG1}
\small{G_1=
\begin{bmatrix}
0 & 0 & 0 & 0 & 0 & 0 &0  & 0 & 0 &0 \\ 
G_{2,1} & 0 & 0 & 0 & 0 & G_{2,6} & 0 & G_{2,8} & 0 &0 \\ 
0 & G_{3,2} & 0 & G_{3,4} & G_{3,5} & 0 & 0 & 0 & G_{3,9} &0 \\ 
0 & G_{4,2} & G_{4,3} & 0 & 0 & G_{4,6} & 0 & G_{4,8} & 0 & 0\\ 
G_{5,1} & 0 & 0 & G_{5,4} & 0 & G_{5,6} & 0 & 0 & 0 & 0\\ 
0 & 0 & 0 & G_{6,4} & G_{6,5} & 0 & 0 & 0 & 0 & 0\\ 
0 & 0 & 0 & 0 & 0 & 0 & 0 & G_{7,8} & 0 &0 \\ 
0& 0 &  0& 0 & G_{8,5} & 0 & G_{8,7} & 0 &0  & 0\\ 
0& G_{9,2} & 0 & 0 & 0 & G_{9,6} & 0 & G_{9,8} & 0 & G_{9,10} \\ 
0 & 0 & 0 & 0 & 0 & 0 & 0 & G_{10,8} & G_{10,9} & 0
\end{bmatrix}}
\end{equation*}

\begin{equation*}
\label{eqG2}
\small{G_2=
\begin{bmatrix}
0 & 0 & 0 & 0 & 0 & 0 & 0 & 0 & 0 & 0\\ 
0 & 0 & 0 & 0 & 0 & 0 & 0 & 0 & 0 & 0\\ 
0 & 0 & 0 & 0 & 0 & 0 & 0 & 0 & 0 & 0\\ 
0 & 0 & 0 & 0 & 0 & 0 & 0 & 0 & 0 & 0\\ 
0 & 0 & 0 & 0 & 0 & 0 & 0 & 0 & 0 & 0\\  
0 & 0 & 0 & 0 & 0 & 0 & 0 & 0 & 0 & 0\\ 
0 & G_{7,12} & 0 & G_{7,14} & 0 & 0 & 0 & 0 & 0 & 0\\ 
0&  G_{8,12}& G_{8,13} & 0 & 0 & 0 & 0 & 0 & 0 & 0\\ 
0 & G_{9,12} & 0 & 0 & 0 & 0 & 0 & 0 &0  &0 \\ 
0 & G_{10,12} & 0 & 0 & 0 &  0&  0&  0& 0 &0 
\end{bmatrix}}
\end{equation*}

\begin{equation*}
\label{eqG3}
\small{G_3=
\begin{bmatrix}
0 & 0 & 0 & 0 & 0 &0  & 0 & 0 & 0 &G_{11,10} \\ 
0 & 0 & 0 & 0 & 0 &0  & 0 & 0 & 0 &G_{12,10} \\ 
0&0  & 0 & 0 & 0 &  0&0  & G_{13,8} & 0 & 0\\ 
0 & 0 & 0 &0  & 0 &0  & 0 & 0 & 0 &0 \\ 
0 & 0 & 0 &0  & 0 &0  & 0 & 0 & 0 &0 \\ 
0 & 0 & 0 &0  & 0 &0  & 0 & 0 & 0 &0 \\ 
0 & 0 & 0 &0  & 0 &0  & 0 & 0 & 0 &0 \\ 
0 & 0 & 0 & 0 & 0 & 0 &0  & 0 & 0 &G_{18,10} \\ 
0 & 0 & 0 &0  & 0 &0  & 0 & 0 & 0 &0 \\ 
0 & 0 & 0 &0  & 0 &0  & 0 & 0 & 0 &0 \\ 
\end{bmatrix}}
\end{equation*}

\begin{equation*}
\label{eqG4}
\small{G_4=
\begin{bmatrix}
0 & G_{11,12} & 0 & 0 & 0 & G_{11,16} & 0 & 0 &0  &0 \\ 
G_{12,11} & 0 & 0 & G_{12,14} & 0 & 0 & 0 & G_{12,18} & 0 & 0\\ 
G_{13,11}& 0 &  0& G_{13,14} &0  & 0 & 0 & 0 &0  & 0\\ 
0& 0 & G_{14,13} & 0 & G_{14,15} & 0 &0  &0  &0  &0 \\ 
0& 0 & 0 & 0 & 0 & G_{15,16} & 0 & G_{15,18} & 0 & 0\\ 
0 & 0 & 0 & G_{16,13} & 0 & 0 & 0 & 0 & 0 & 0\\ 
0 & 0 & 0 & 0 & 0 &  G_{17,16}& 0 & G_{17,18} & G_{17,19} & 0\\ 
0& 0 & 0 & 0 & 0 & 0 & G_{18,17} & 0 & 0 & 0\\ 
0 & G_{19,12} & 0 & G_{19,14} & 0 & 0 & 0 & G_{19,18} & 0 & 0\\ 
0 & G_{20,12} &  G_{20,13}& 0 & 0 & 0 & 0 &0  & 0 &0 
\end{bmatrix}}
\end{equation*}
The transfer functions are described below:

\begin{eqnarray*}
	G_{2,1} = -1.1576491 \times 10^{-01}+4.2048459 \times 10^{-02} z^{-1}\\
	G_{2,6} = -4.9391907 \times 10^{-01}+2.3094301 \times 10^{-01} q^{-1}\\
	G_{2,8} = -3.8295603 \times 10^{-01}+3.7364537 \times 10^{-01} q^{-1}\\
	G_{3,2} = -2.3501597 \times 10^{-01}+2.2411979 \times 10^{-01} q^{-1}\\
	G_{3,4} = -0.3q^{-1}+0.8q^{-2}\\
	G_{3,5} = -0.5q^{-1}\\
	G_{3,9} = -1.5484356 \times 10^{-01}+3.5947903 \times 10^{-01} q^{-1}\\
	G_{4,2} = -3.4361929 \times 10^{-01}+2.7664996 \times 10^{-01} q^{-1}\\
	G_{4,3} = q^{-1}\\
	G_{4,6} = -4.4565148 \times 10^{-02}+3.1267256 \times 10^{-02} q^{-1}\\
	G_{4,8} = -3.0217221 \times 10^{-02}+4.9084253 \times 10^{-01} q^{-1}\\
	G_{5,1} = -4.4755747 \times 10^{-01}+1.5153359 \times 10^{-01} q^{-1}\\
	G_{5,4} = 0.5q^{-1}\\
	G_{5,6} = -1.8258082 \times 10^{-02}+2.5655941 \times 10^{-02} q^{-1}\\
	G_{6,4} = -4.0083967 \times 10^{-02}+2.3831631 \times 10^{-02} q^{-1}\\
	G_{6,5} = -4.9526830 \times 10^{-02}+1.8655891 \times 10^{-02} q^{-1}\\
	G_{7,8} = -4.2353188 \times 10^{-02}+1.7016841 \times 10^{-03} q^{-1}\\
	G_{7,12} = -3.8831215 \times 10^{-01}+1.6625282 \times 10^{-01} q^{-1}\\
	G_{7,14} = -1.3013545 \times 10^{-01}+3.2468616 \times 10^{-01} q^{-1}\\
	G_{8,5} = -4.8312501 \times 10^{-01}+2.9208833 \times 10^{-01} q^{-1}\\
	G_{8,7} = -4.3341455 \times 10^{-02}+2.6095021 \times 10^{-02} q^{-1}\\
	G_{8,12} = -2.0610019 \times 10^{-01}+2.6998910 \times 10^{-01} q^{-1}\\
	G_{8,13} = -1.4342078 \times 10^{-02}+3.4009137 \times 10^{-02} q^{-1}\\
	G_{9,2} = -2.0348115 \times 10^{-01}+6.7364494 \times 10^{-02} q^{-1}\\
	G_{9,6} = -2.9096829 \times 10^{-01}+6.1878182 \times 10^{-02} q^{-1}\\
	G_{9,8} = -4.7096177 \times 10^{-01}+1.6149849 \times 10^{-01} q^{-1}\\
	G_{9,10} = -3.6050395 \times 10^{-02}+4.1517977 \times 10^{-02} q^{-1}\\
	G_{9,12} = -3.1296376 \times 10^{-02}+1.1860562 \times 10^{-01} q^{-1}\\
	G_{10,8} = -3.0338765 \times 10^{-01}+4.1470173 \times 10^{-01} q^{-1}\\
	G_{10,9} = -3.4408597 \times 10^{-02}+2.3732489 \times 10^{-03} q^{-1}\\
	G_{10,12} = -3.4207005 \times 10^{-02}+4.4904179 \times 10^{-02} q^{-1}\\
	G_{11,10} = \frac{2.4710993 \times 10^{-01} q^{-1}}{1-5.0578013 \times 10^{-01}q^{-1}}\\
	G_{11,12} = \frac{2.4512609 \times 10^{-02} q^{-1}}{1-5.0974782 \times 10^{-01}q^{-1}}\\
	G_{11,16} = \frac{2.3010071 \times 10^{-01} q^{-1}}{1-5.3979857 \times 10^{-01}q^{-1}}\\
	G_{12,10} = \frac{2.0528463 \times 10^{-02} q^{-1}}{1-5.8943074 \times 10^{-01}q^{-1}}\\
	G_{12,11} = \frac{2.1646986 \times 10^{-02} q^{-1}}{1-5.6706027 \times 10^{-01}q^{-1}}\\
	G_{12,14} = \frac{2.0877819 \times 10^{-01} q^{-1}}{1-5.8244362 \times 10^{-01}q^{-1}}\\
	G_{12,18} = \frac{2.0657294 \times 10^{-01} q^{-1}}{1-5.8685411 \times 10^{-01}q^{-1}}\\
\end{eqnarray*}
\begin{eqnarray*}
	G_{13,8} = \frac{2.1848002 \times 10^{-02} q^{-1}}{1-5.6303996 \times 10^{-01}q^{-1}}\\
	G_{13,11} = \frac{2.2137643 \times 10^{-01} q^{-1}}{1-5.5724714 \times 10^{-01}q^{-1}}\\
	G_{13,14} = \frac{2.2709971 \times 10^{-02} q^{-1}}{1-5.4580058 \times 10^{-01}q^{-1}}\\
	G_{14,13} = \frac{2.2787928 \times 10^{-02} q^{-1}}{1-5.4424144 \times 10^{-01}q^{-1}}\\
	G_{14,15} = \frac{2.4571974 \times 10^{-01} q^{-1}}{1-5.0856052 \times 10^{-01}q^{-1}}\\
	G_{15,16} = \frac{2.4854075 \times 10^{-01} q^{-1}}{1-5.0291850 \times 10^{-01}q^{-1}}\\
	G_{15,18} = \frac{2.0964010 \times 10^{-01} q^{-1}}{1-5.8071980 \times 10^{-01}q^{-1}}\\
	G_{16,13} = \frac{2.1627442 \times 10^{-01} q^{-1}}{1-5.6745117 \times 10^{-01}q^{-1}}\\
	G_{17,16} = \frac{2.3606224 \times 10^{-01} q^{-1}}{1-5.2787553 \times 10^{-01}q^{-1}}\\
	G_{17,18} = \frac{2.4035419 \times 10^{-02} q^{-1}}{1-5.1929163 \times 10^{-01}q^{-1}}\\
	G_{17,19} = \frac{2.3030840 \times 10^{-01} q^{-1}}{1-5.3938319 \times 10^{-01}q^{-1}}\\
	G_{18,10} = \frac{2.1838053 \times 10^{-01} q^{-1}}{1-5.6323893 \times 10^{-01}q^{-1}}\\
	G_{18,17} = \frac{2.3869253 \times 10^{-02} q^{-1}}{1-5.2261494 \times 10^{-01}q^{-1}}\\
	G_{19,12} = \frac{2.4800810 \times 10^{-01} q^{-1}}{1-5.0398380 \times 10^{-01}q^{-1}}\\
	G_{19,14} = \frac{2.4554410 \times 10^{-01} q^{-1}}{1-5.0891181 \times 10^{-01}q^{-1}}\\
	G_{19,18} = \frac{2.3382918 \times 10^{-01} q^{-1}}{1-5.3234164 \times 10^{-01}q^{-1}}\\
	G_{20,12} = \frac{2.1965134 \times 10^{-01} q^{-1}}{1-5.6069731 \times 10^{-01}q^{-1}}\\
	G_{20,13} = \frac{2.2859570 \times 10^{-01} q^{-1}}{1-5.4280860 \times 10^{-01}q^{-1}}\\
\end{eqnarray*}

\end{document}